\documentclass[prb,twocolumn,amsmath,showpacs,preprintnumbers,superscriptaddress,amssymb,space,data]{revtex4}
\usepackage{amssymb}
\usepackage{makeidx}
\usepackage{amsmath}
\usepackage{dcolumn}
\usepackage{bm}
\usepackage{graphicx}
\usepackage{amsfonts}

\begin{document}
\title{Amplitude spectroscopy of two coupled qubits}

\author{A. M. Satanin}
\email{sarkady@mail.ru}
\author{M. V. Denisenko}

\affiliation{Nizhny Novgorod State University, 23 Gagarin Ave., 603950, Nizhny Novgorod, Russia}

\author{Sahel Ashhab}

\author{Franco Nori}

\affiliation{Advanced Science Institute, RIKEN, Wako-shi, Saitama, 351-0198, Japan}
\affiliation{Department of Physics, University of Michigan, Ann Arbor, MI 48109-1040, USA}

\begin{abstract}

We study the effect of a time-dependent driving field with a large amplitude on a system composed of two coupled qubits (two-level systems). Using the rotating wave approximation (RWA) makes it possible to find simple conditions for resonant excitation of the four-level system. We find that the resonance conditions include the coupling strength between the qubits. Numerical simulations confirm the qualitative conclusions following from the RWA. To reveal the peculiarities of resonant transitions caused by the quasi-level motion and crossing in a periodic driving field, we use Floquet states, which determine the precise intermediate states of the system. Calculating the quasi-energy states of the multi-level system makes it possible to find the transition probabilities and build interference patterns for the transition probabilities. The interference patterns demonstrate the possibility of obtaining various pieces of information about the qubits, since the positions of transition-probability maxima depend on various system parameters, including the coupling strength between the qubits.


\end{abstract}

\maketitle

\section{Introduction}
Recently much attention has been focused on the spectroscopy of Josephson junction-superconducting circuits with a weak link which can be considered as ``macroscopic atoms'' with sizes of the order of tens or hundreds of micrometers \cite{Nori1,Nori3}. Single Josephson-junction qubits are characterized by relatively long relaxation times (tens of microseconds) which allows to consider them as one of the most promising elements for the realization of quantum information processors \cite{Zagoskin}. Spectroscopic investigations of artificial ``Josephson atoms'' are carried out at sufficiently low temperatures in  the micro-wave and millimeter-wave regions, since the spectral lines of Josephson junctions are located in that spectral region. However, practical measurements are not simple because stable tunable micro-wave sources are not easy to produce in this range. Measurement difficulties are connected with the frequency dependence of the dispersion and damping, as well as with strict requirements to impedance control tolerances which limit the application of  broadband spectroscopy.

In this regard several groups have used amplitude spectroscopy
\cite{Oliver,Berns1,Sillanpa,Berns2,Ruder}
which obtains information by means of the response function ``sweep'' over the signal amplitude and some control parameter (an applied magnetic flux or a bias). This method may be applied to systems with crossing energy levels between which the transitions can be realized by changing external parameters. The frequency of such a driving field can be orders of magnitude lower than the distance between levels. This means that the system evolves adiabatically, except for the immediate vicinity of quasi-crossing levels, where Landau-Zener quantum coherent transitions and St\"{u}ckelberg interference can be observed \cite{Landau,Zener,Stuckelberg,Majorana} (see Ref.~\onlinecite{Shevchenko1} for an overview). The main advantage of this type of spectroscopy is that the system is investigated in wide ranges of the amplitude change. Thus, in  alternating fields, multiphoton processes and  Landau-Zener transitions, also observed earlier in Ref.~\onlinecite{Multiphoton}, take place.

For a driven two-level system, drastic effects on the tunneling rate arise from quasi-energy crossing and anticrossing \cite{Autler,Shirley}. At certain amplitudes of the driving field, dynamical localization and trapping of the system into a non-linear resonance can take place \cite{Scully,Grossmann,Agarwal,Ian,Gawryluk,Zhang,Tuorila,Wubs,Childress}. As the parameters are changed when the level-approaching and level-crossing take place, the effects of band-to-band tunneling (Landau-Zener transitions) can occur. In the language of wave functions, an interference of different zone states, predicted by St\"{u}ckelberg \cite{Stuckelberg} is possible. As applied to qubits these effects have lately been discussed in numerous works \cite{Ashhab,Son,Sun,Sun2,Plotz,Wang,Hausinger,Xu,Du,Ferron,Denisenko,Hijii,Ditzhuijzen,Wang2,Chotorlishvili,Gasparinetti}.

Coupled qubits have also been created (see, e.g., Ref.~\onlinecite{Berkley,Izmalkov,
Majer,Steffen,van der Ploeg,Plantenberg, Izmalkov2,Groot,Shevchenko3}). In these works the basic parameters of qubits and coupling constants have been measured and also some relaxation characteristics of coupled qubits have been studied. Rabi-spectroscopy of two coupled qubits both experimentally and theoretically have been investigated
in publications \cite{Majer,Steffen,Plantenberg,Groot,Shevchenko3}. Recently different schemes of controlled coupling between two or more qubits have been proposed \cite{You, Ashhab2, Ashhab3,Izmalkov2}. However, at present there are no studies of the way the coupled qubits behave in strong fields. Meanwhile, the extension of the amplitude spectroscopy method makes it possible to give much information about coupled multi-qubit clusters.

The goal of this work is to describe quantum-mechanical phenomena in a system of coupled qubits from the point of view of quasi-energy states at different parameters of multi-level systems. It is possible to control the magnetic fluxes (biases) which penetrate the circuits and we will take these bias parameters to be dependent on time \cite{Berkley,Izmalkov,
Majer,Steffen,
van der Ploeg,Plantenberg,Izmalkov2,Groot,Shevchenko3}. Although generally, for spectroscopic investigations, the response dependence on the frequency is studied, we will focus our attention here on the response dependence on the signal amplitude and the control parameters. Our approach differs from the one used in Ref.~\onlinecite{Temchenko,Storcz}, where the density matrix equation was used to determine the steady state populations of coupled qubits. We assume here that the qubits are relevant for quantum information processing only in the case when they have negligible dissipation. In this case, to analyze the dynamics of the system it is most natural to proceed directly from the Schr\"{o}dinger equation, which allows us to understand the dynamics and to identify features of the evolution of systems in strong alternating fields.

First, we shall investigate the nonlinear time dynamics of the coupled qubits by using the RWA In this approximation the system exhibits generalized Rabi resonances where the role of the coupling parameter may be investigated. Secondly, for high-field amplitude excitations we shall apply the quasi-energy representation to understand the influence of the driving field on the transition probabilities and the population of the energy levels. Using the RWA  and numerical calculations of quasi-energy levels as a function of the driving field we will be able to demonstrate that the effect of the quasi-energy avoided crossing leads to drastically increased transition probabilities between the qubits steady states. Finally, we shall develop a numerical method for calculating the transition probabilities in the quasi-energy representation and build interference patterns for the transition probabilities. The quasi-energy basis allows us to analyze the influence of phase fluctuations on the observable effects that have not previously been studied in previous works. As we demonstrate in the following, the peaks of the transition probabilities between the directly coupled states shift if the inter-qubit coupling changes, the indirectly coupled states the peak positions are not affected by the inter qubit coupling. This effect can be observed in experiments using the technique of amplitude spectroscopy. It will be demonstrated that Landau-Zener-St\"{u}ckelberg interferometry or amplitude spectroscopy may be considered as a tool to obtain the coupling parameter by seeing the shift of the peak of the resonances (the population maxima).

\section{EQUATION OF MOTION OF COUPLED QUBITS}

The main features of coupled qubits system behavior can be understood in the framework of the Hamiltonian:
%
%
\begin{equation}
H=-\frac{1}{2}\left(
               \begin{smallmatrix}
               \epsilon_{1}+\epsilon_{2}+J & \Delta_{2} & \Delta_{1} & 0\\
               \Delta_{2} & \epsilon_{1}-\epsilon_{2}-J & 0 & \Delta_{1}\\
               \Delta_{1} & 0 & -\epsilon_{1}+\epsilon_{2}-J & \Delta_{2}\\
               0 & \Delta_{1} & \Delta_{2} & -(\epsilon_{1}+\epsilon_{2})+J
              \end{smallmatrix}
              \right),\label{1}%
\end{equation}
where $\epsilon_{i}$ is the control parameter of qubit $i$ ($i=1,\,2$), $\Delta_{i}$ is the corresponding tunneling matrix element, and the parameter $J$ quantifies the strength of the interaction between the qubits.
The form of the Hamiltonian differs from \cite{Majer,Storcz,Temchenko}
only by a simple redefinition of parameters.

Near the half-flux quantum point, each flux qubit experiences a double-well potential and the tunneling energy through the potential barrier separating the wells becomes $\Delta_{i}$. The wells correspond to currents of magnitude $I_{i}$ circulating in opposite directions along the loop, and the above Hamiltonian is actually written in this circulating current basis. Following Ref. \onlinecite{Izmalkov,Majer}, in a constant field the control parameters $\epsilon_{i}$ can be expressed in terms of the bias $f_{i}=\Phi^{\rm{ext}}_{i}/\Phi_{0}$  ($\Phi^{\rm{ext}}_{i}$ is the flux threading the qubit loop (magnetic flux), penetrating circuit $i$, $\Phi_{0}$    is the flux quantum) by the relation
\begin{equation}
\epsilon_{i}=\epsilon^{0}_{i}\left(f_{i}-\frac{1}{2}\right),\label{2}%
\end{equation}
where $\epsilon^{0}_{i}=2|I_{i}|\Phi_{0}$. The parameters $\epsilon_{i}$ and $\Delta_{i}$ determine the spectrum of the uncoupled qubits ($J=0$):
$
E_{i}=\pm\frac{1}{2}\sqrt{\epsilon^{2}_{i}+\Delta^{2}_{i}}.
$
The ferromagnetic/antiferromagnetic interaction between the qubits is characterized by the coupling strength $J=\pm|J|$.  With the help of an additional superconducting circuit it is possible to realize ferromagnetic as well as antiferromagnetic interactions between the qubits \cite{Izmalkov2}.  For a planar circuit the antiferromagnetic interaction is determined by the expression $\frac{|J|}{2}=M_{12}I_{1}I_{2}$, where $M_{12}$ is the mutual inductance.

The state of the system can be represented by four amplitudes $C_{\alpha}(t)$, $\alpha=1,... ,4$, so that $|\Psi\rangle=\sum C_{\alpha}(t)|\alpha\rangle$, where $|\alpha\rangle$ is the basis of the time-independent Hamiltonian Eq.~(\ref{1}). The spectrum $E_{\alpha}$ and eigenvectors $|\alpha\rangle$ of the Hamiltonian ({\ref{1}}) are not difficult to find.

To study the time-dependent evolution of the coupled qubits we use the eigenstates of the Hamiltonian Eq.~(\ref{1}) as the basis, since expanding in this basis is a well controlled procedure.
Let us now consider the case when the control parameters $\epsilon_{1,2}$ are time-dependent.
For the case of coupled qubits, we introduce a driving field of the form
\begin{equation}
\epsilon_{1}(t)=\epsilon_{10}+A_{1}\cos(\omega_{1}t+\theta_1),\:\:\:\:\: \epsilon_{2}(t)=\epsilon_{20}+A_{2}\cos(\omega_{2}t+\theta_2).\label{4}%
\end{equation}
For simplicity, we will only discuss the case when driving fields of only one frequency $\omega=\omega_{1}=\omega_{2}$ are applied to the system and the two fields have the same phase shift $\theta=\theta_{1}=\theta_{2}$.
In this paper, we also assume that the system is subject to a sequence of synchronized pulses of alternating fields whose duration is much longer than the period of the field. At the same time, we take into account the fluctuations in the arrival times of pulses and their durations against a fixed period of the field \cite{Shirley}.

We will solve the time-dependent Schr\"{o}dinger equation to determine the resonant conditions of the qubits,
%
%
%
%
\begin{equation}
i\hbar\frac{\partial}{\partial t}|\Psi(t)\rangle=H(t)|\Psi(t)\rangle.\label{5}%
\end{equation}
We perform the canonical transformation:
\begin{equation}
|\Psi(t)\rangle=U(t)|\overline{\Psi}(t)\rangle\, \label{6}%
\end{equation}
where the unitary matrix
$
U(t)=\exp{[iS(t)/2\hbar]},
$
with
\begin{equation}
S(t)=\phi_{1}(t)\left(\begin{smallmatrix}
                 I & 0\\
                 0 & -I
                 \end{smallmatrix} \right)+
                 \phi_{2}(t)\left(\begin{smallmatrix}
                            \sigma_{z} & 0 \\
                            0 & \sigma_{z}
                            \end{smallmatrix}\right)\\+Jt\left(\begin{smallmatrix}
                                                            \sigma_{z} & 0\\
                                                            0 & -\sigma_{z}
                                                            \end{smallmatrix}\right),\label{8}%
\end{equation}
and phases $\phi_{1,2}(t)=\epsilon_{(1, 2)0}t+\frac{A_{1,2}}{\hbar\omega}\sin{\omega t}$. The transformed Hamiltonian $\overline{H}$ has the following form
%
%
%
%
\begin{widetext}
\begin{multline}
\overline{H}(t)=-\frac{\Delta_{1}}{2}\sum^{\infty}_{n=
                -\infty}{J_{n}\left(\frac{A_{1}}{\hbar\omega}\right)}\times
                                                     \left(
                                                    \begin{smallmatrix}
                                                     0 & 0 & e^{-i((\epsilon_{10}+J)/\hbar+n\omega)t} & 0\\
                                                     0 & 0 & 0 & e^{-i((\epsilon_{10}-J)/\hbar+n\omega)t}\\
                                                     e^{i((\epsilon_{10}+J)/\hbar+n\omega)t} & 0 & 0 & 0 \\
                                                     0 & e^{i((\epsilon_{10}-J)/\hbar+n\omega)t} & 0 & 0
                                                     \end{smallmatrix}
                                                     \right)
                                                     \\-\frac{\Delta_{2}}{2}\sum^{\infty}_{n=
                -\infty}{J_{n}\left(\frac{A_{2}}{\hbar\omega}\right)}\times\left(
                                                    \begin{smallmatrix}
                                                     0 & e^{-i((\epsilon_{20}+J)/\hbar+n\omega)t}& 0 & 0\\
                                                     e^{i((\epsilon_{20}+J)/\hbar+n\omega)t}& 0 & 0 & 0 \\
                                                     0 & 0 & 0 & e^{-i((\epsilon_{20}-J)/\hbar+n\omega)t}\\
                                                     0 & 0 & e^{i((\epsilon_{20}-J)/\hbar+n\omega)t} & 0
                                                     \end{smallmatrix}
                                                     \right),\label{9}%
\end{multline}
\end{widetext}
where the following relation for Bessel functions was used
\begin{equation}
\exp{\left(i\frac{A}{\hbar\omega}\sin{\omega t}\right)}=
\sum_{n}{J_{n}\left(\frac{A}{\hbar\omega}\right)\exp{\left(in\omega t\right)}}.
\end{equation}
From Eq.~({\ref{9}}) it follows that the resonance conditions are given by $\epsilon_{10}\pm J+n\hbar\omega\approx0$, $\epsilon_{20}\pm J+n\hbar\omega\approx0$, and à \textit{population trapping} is controlled by the two conditions $J_{n}(\frac{A_{1}}{\hbar\omega})=0$ and $J_{n}(\frac{A_{2}}{\hbar\omega})=0$. It is evident that in this case the resonance conditions depend on the coupling constant.
In the RWA in the Hamiltonian Eq.~({\ref{9}}) fast oscillating components can be neglected with the exception of those for which the resonance conditions are satisfied. Then the Hamiltonian describing the slow dynamics will have the simple matrix form which we can find, in general, the four quasi-energies.

It should be noted that the obtained results are valid in the framework of the RWA \cite{Scully} and cannot describe the system dynamics at an arbitrary amplitude time-dependent field. To leave the framework of the RWA limitations we will apply the numerical solution  of the Schr\"{o}dinger equation in the next section. Recent studies beyond the RWA can be found in Refs.~[\onlinecite{Ashhab, Son, Wang, Hausinger, Xu}].
\section{QUASI-ENERGIES AND TRANSITION AMPLITUDES IN A STRONG DRIVING FIELD}
To obtain results for high-field amplitudes a quasi-energy representation is used. This representation gives the precise intermediate system state in a periodically-driven field with an arbitrary amplitude and allows to detect the peculiarities of resonant transitions caused by the motion and crossing of quasi-levels when the field changes.

\subsection{Quasi-energies of multi-level systems}

Let us consider the Hamiltonian of a multi-level system and let us take it to be periodic with period $T = 2\pi/\omega$
\begin{equation}
H(t)=H(t+T). \label{10}
\end{equation}
According to Floquet's theorem, the general solution of the Schr\"{o}dinger equation can be decomposed into the complete set of functions
\begin{equation}
|\Psi_{k}(t)\rangle=|\Phi_{k}(t)\rangle e^{-iQ_{k}t/\hbar},\quad
|\Phi_{k}(t+T)\rangle=|\Phi_{k}(t)\rangle, \label{11}%
\end{equation}
where the functions $|\Phi_{k}(t)\rangle$ are the solutions of the equation
\begin{equation}
\left(H(t)-i\hbar\frac{\partial}{\partial t}\right)|\Phi_{k}(t)\rangle=Q_{k}|\Phi_{k}(t)\rangle,\label{12}
\end{equation}
and the real parameter $Q_{k}$ is called the quasi-energy \cite{Shirley,quasienergy}($k$ is the quantum number determining the quasi-energy).

The quasi-energies $Q_{k}$ and eigenfunctions $|\Phi_{k}(0)\rangle$ at the initial moment of time (which may be chosen arbitrarily \cite{Shirley}) are found by the solution
\begin{equation}
F(T)|\Phi_{k}(0)\rangle=e^{-iQ_{k}T/\hbar}|\Phi_{k}(0)\rangle, \label{13}
\end{equation}
where $F(T)=\hat{P}\exp(-i\int^{T}_{0}{H(t)dt}/\hbar)$, $\hat{P}$ is the chronological ordering operator. The value of the functions $|\Phi_{k}(t)\rangle$ at any moment of time are obtained from the equation (\ref{12}). Since quasi-energies are not uniquely defined $Q^{'}_{k}=Q_{k}+n\hbar\omega$, we will depict them in the first ``Brillouin'' zone ($0<Q_{k}<\hbar\omega$).

Expanding the periodic functions $|\Phi_{k}(t)\rangle$ in Fourier series \cite{Autler,Shirley,quasienergy} can be used to find the quasi-energies. The coefficients of the Fourier series in turn satisfy an infinite-dimensional system of linear equations which is approximately solved by a finite-dimensional approximation. In this work the form of the functions $Q_{k}$ is found numerically. First, we do not need to work with large-size sub-matrices; secondly, this approach allows us to obtain a controllable approximate solution.

An arbitrary wave function may be expanded in the complete Floquet basis
\begin{equation}
|\Psi(t)\rangle=\sum_{k}{c_{k}}|\Phi_{k}(t)\rangle e^{-iQ_{k}t/\hbar}, \label{14}
\end{equation}
where the coefficients $c_{k}$ are defined by the initial wave function: $c_{k}=\langle\Phi_{k}(0)|\Psi(0)\rangle$. So the Floquet time-evolutional operator can be found from Eq. (\ref{14}):
\begin{equation}
F(t,0)=\sum{|\Phi_{k}(t)\rangle e^{-iQ_{k}t/\hbar}\langle\Phi_{k}(0)|}.\label{15}%
\end{equation}

Let us take the system to be initially in the state $|\alpha\rangle$, which is a steady state of the time-independent Hamiltonian Eq.~(\ref{1}). Let us also suppose that the electromagnetic pulse has an unknown phase. The transition probability into the excited state $|\beta\rangle$ of the Hamiltonian Eq.~(\ref{1}), averaged over the relative phase, is described by the following expression:
\begin{equation}
P_{\alpha\rightarrow\beta}(t)=\sum_{k,l}
e^{-i(Q_{k}-Q_{l})t/\hbar}
\sum_{n}M_{k}^{(n)}(t)M_{l}^{*(n)}(t),\label{16}%
\end{equation}
where
\begin{equation}
M_{k}^{(n)}(t)=\frac{1}{T}\int^{T}_{0}{e^{-in\omega \tau}
\langle\beta|\Phi_{k}(\tau+t)\rangle\langle\Phi_{k}(\tau)|\alpha\rangle
d\tau}.\label{17}%
\end{equation}
Notice that the sum with respect to $n$ appears in Eq.~(\ref{16}) because the Fourier expansion of the Floquet states has been used in the intermediate manipulations.

The expression Eq.~(\ref{15}) manifests that in a strong field the system evolution occurs through the intermediate quasi-energy states of qubits.
It may be shown that the transition probability Eq.~(\ref{16}) in general contains strongly oscillating-in-time terms which may be be canceled when the time interval $t$ is long enough.
The exception is the contribution of the states with almost equal quasi-energies.
After averaging the expansion for the probability Eq.~(\ref{16}) by the time interval  $t$  we find
\begin{equation}
\overline{P}_{\alpha\rightarrow\beta} = \sum_{k}\sum_{n, n^{'}}\left |\langle\beta|\Phi^{(n-n^{'})}_{k}\rangle\langle\Phi^{(n)}_{k}|\alpha\rangle\right |^{2} ,\label{18}
\end{equation}
where the Fourier components are defined by the relation
\begin{equation}
|\Phi^{(n)}_{k}\rangle=\frac{1}{T}\int^{T}_{0}\!\!{e^{in\omega t}\;|\Phi_{k}(t)\rangle dt}.\label{19}%
\end{equation}
The transition probabilities for different harmonics can be calculated according to Eq.~(\ref{18}). To do that we solve numerically Eq.~(\ref{12}) and take the Fourier components according to Eq.~(\ref{19}).

\subsection{Numerical results for coupled qubits in a strong driving field}

We now present numerical results of the coupled qubits response in a strong driving field. We will use the language of quasi-energies crossing which depend on the system parameters.
It is well known that when the amplitude of the driving field and control parameter change, the quasi-energies of different symmetry classes may cross but if they are of the same symmetry class they form an anticrossing. As a result the transition amplitudes may change drastically for such
parameters \cite{Autler,Shirley,quasienergy,Grifoni}.
Special attention will be paid to the dependence of the level populations on the interaction parameter. As was recently shown \cite{van der Ploeg}, the interaction parameter can be varied over a wide range by using an intermediate coupler which, for instance, may be an additional Josephson loop placed between the two main qubit loops. So, we are going to investigate here the behavior of the level populations as a function of the coupling parameter of the qubits.

\begin{figure}[t]
\begin{center}
    \includegraphics[width=8.5cm,height=12cm]{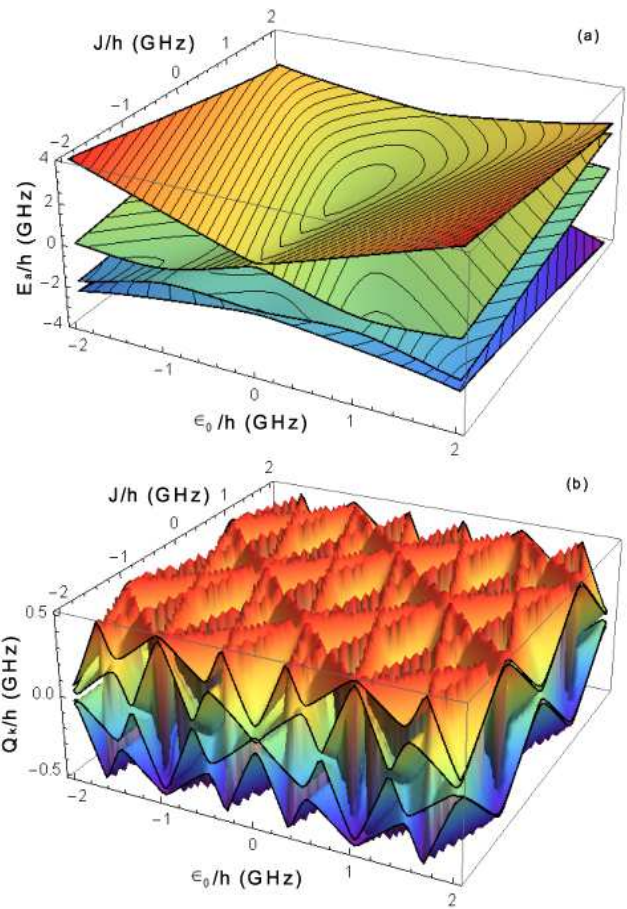}
\end{center}
\caption{\label{fig1}(Color online). (a) Energies $E_\alpha$ of the Hamiltonian (\ref{1}), and (b) the quasi-energies $Q_k$ as  functions of the control parameters $\epsilon_0 = \epsilon_{20} = 2 \epsilon_{10}$ and  the coupling parameter $J$. We used the qubit parameters: $\Delta_{2}/h = 1.5 \Delta_{1}/h = 0.45$ GHz, $\omega/2\pi = 1$ GHz, and $A_{2}/h = 2A_{1}/h = 7$ GHz.}
\end{figure}

First we shall depict a 3D plot of the qubit energy dependence on the control parameter and the coupling parameter. Figure~\ref{fig1}(a) shows the energy surfaces for the time-independent Hamiltonian Eq.~(\ref{1}) (when $A=0$). Figure~\ref{fig1}(b) shows the transformation of the dispersion surfaces to quasi-energy surfaces when the time-dependent field is applied to the qubits. In order to understand what quasi-energies cross, we have depicted some of the characteristic cross sections of the quasi-energy surface in Fig.~\ref{fig1}(b).
%
%
%
%
%
\begin{figure*}[t]
\begin{center}
    \includegraphics[width=12cm,height=7cm]{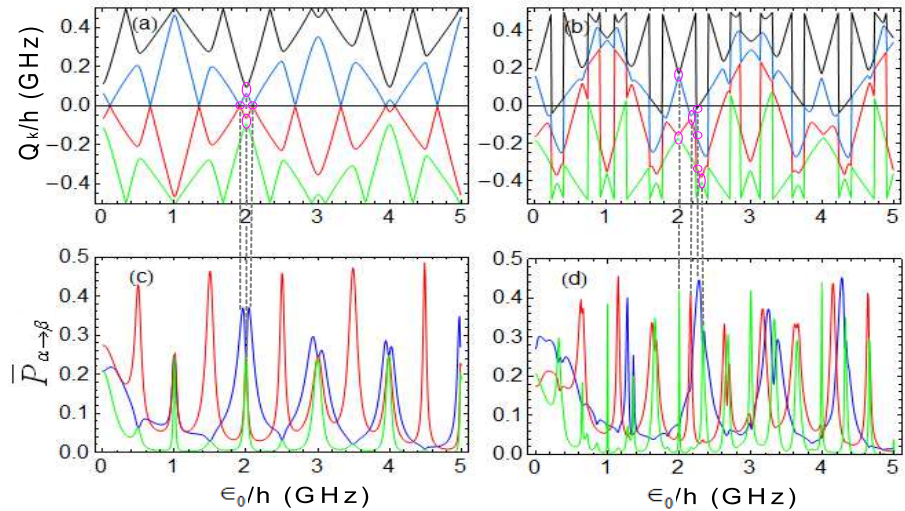}
\end{center}
\caption{\label{fig2}(Color online). (a, b) The quasi-energies as functions of the control parameter $\epsilon_{0}$ for an external amplitude $A/h =4$ GHz. The bottom row (c, d) shows the transition probabilities $\overline{P}_{1\rightarrow2}$  (blue), $\overline{P}_{1\rightarrow3}$ (red) and $\overline{P}_{1\rightarrow4}$  (green). The coupling parameter $J=0$ for the left column (a, c), and $J/h = -0.1$ GHz for the right column (b, d) were chosen. The following qubit parameters were used here $\omega/2\pi=1$ GHz, $\Delta_{2}/h=1.5\Delta_{1}/h=0.45$ GHz, $\epsilon_{0} = \epsilon_{20} = 2\epsilon_{10}$, and $A = A_{2}= 2 A_{1}$.
}
\end{figure*}

%
\begin{figure*}
\begin{center}
    \includegraphics[width=12cm,height=7cm]{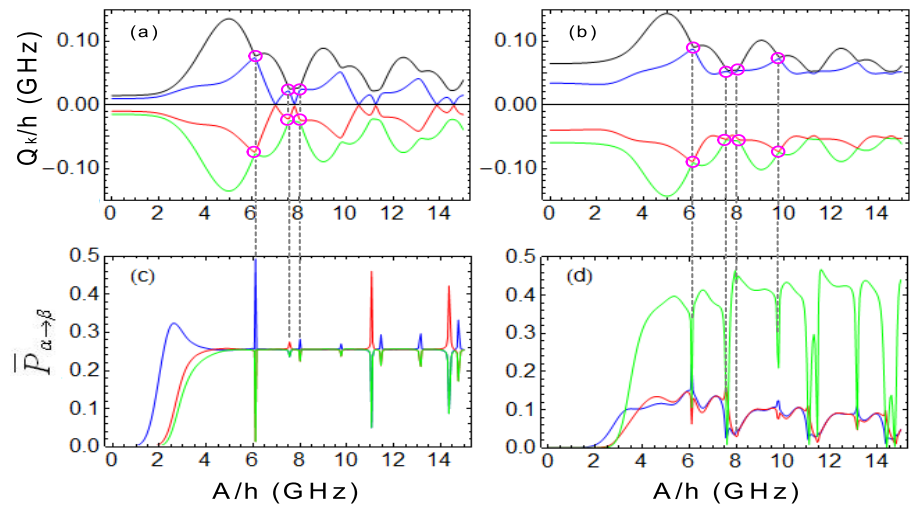}
\end{center}
\caption{\label{fig3}(Color online). The upper row line (a,b) presents quasi-energy versus the field amplitude $A$ (which is also measured in GHz) for coupled qubits; $\epsilon_{0} = \epsilon_{20} = 2\epsilon_{10}$, $\epsilon_{0}/h =4$ GHz. The bottom row (c, d) shows the transition probabilities $\overline{P}_{1\rightarrow2}$  (blue), $\overline{P}_{1\rightarrow3}$ (red) and $\overline{P}_{1\rightarrow4}$ (green). The other parameters and designations are the same as in Fig.~\ref{fig2}.
}
\end{figure*}
%
\begin{figure*}
\begin{center}
      \includegraphics[width=17cm,height=10cm]{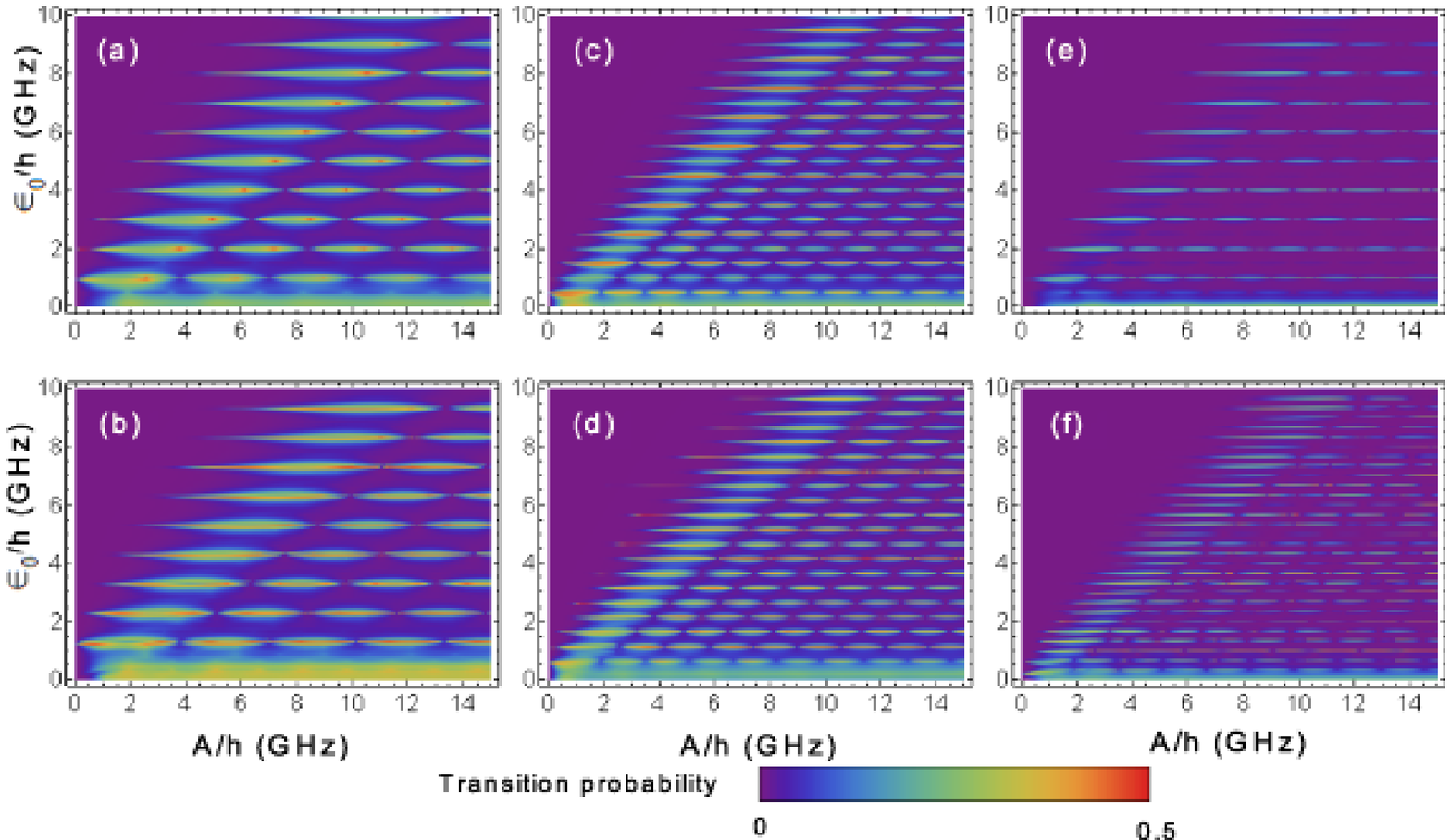}
\end{center}
\caption{\label{fig4}(Color online). The transition probabilities: $\overline{P}_{1\rightarrow2}$ (a, b), $\overline{P}_{1\rightarrow3}$ (c, d), and $\overline{P}_{1\rightarrow4}$ (e, f) of two coupled  qubits ($\Delta_{2}/h = 1.5\Delta_{1}/h = 0.45$ GHz) as functions of the driving amplitude field $A= A_2 = 2 A_1$, ($\omega/2\pi=1$ GHz) and control parameter $\epsilon_0 = \epsilon_{20} = 2 \epsilon_{10}$ for different values of the coupling parameter: (a, c, e) $J=0$ and (b, d, f) $J/h =-0.3$ GHz.}
\end{figure*}
The dependencies of the quasi-energies and transition probabilities on the control parameter, at a given amplitude of the alternating field, were investigated. In Fig.~\ref{fig2}(a, b) the quasi-energies are shown as functions of the control parameter $ \epsilon_0=\epsilon_{20}=\lambda\epsilon_{10}$ for $J=0$ (a) and $J/h=-0.1$ GHz (b), respectively.
In this case, a set of quasi-energy level crossings which produce additional peaks for transition probabilities between the eigenstates of the Hamiltonian Eq.~(\ref{1}) is observed, in Fig.~\ref{fig2}(c, d). Several examples of the quasi-energy level crossings in Fig.~\ref{fig2} and their coincidence with the resonance peaks are shown by the gray vertical dotted lines.

The quasi-energy dependence, in Fig.~\ref{fig2}(b), on the control parameter can be easily understood in the framework of perturbation theory. We shall explain the meaning of the quasi-energy levels formation, which is  shown in Fig.~\ref{fig2}(a). Let us mentally draw a set of lines parallel to the vertical axis at distances $n\hbar\omega$ from each other and then move the fragments of dispersion curves from each line to the first Brillouin zone ($0<Q_{k}<\hbar\omega$). It is shown below that the obtained picture will approximately correspond to the pictures shown in Fig.~\ref{fig2}. As can be seen from Fig.~\ref{fig2}(a) the dependence of quasi-energies on the parameter $\epsilon_{0}$ is very simple at $\epsilon_{0}\gg\Delta$: the quasi-energies behave in accordance with the almost linear laws of dispersion of the uncoupled qubits (defined by $\hbar\omega$ module).
The above explanation also provides a key to understanding the meaning of Fig.~\ref{fig1} (b).
Notice that when $\epsilon_{0}\sim\Delta$, the curvature of the qubits dispersion plays an important role in the formation of the resonance peaks [see Fig.~\ref{fig2}(d)].

Figure~\ref{fig3}(a) shows the dependence of the four quasi-levels of two non-interacting qubits in an alternating field.
In the RWA, the dependence of the quasi-energies on the driving amplitude may be found approximately from the average Hamiltonian defined by Eq.~(\ref{9}). The inclusion of the interaction leads to an effective repulsion of quasi-energy levels [Fig.~\ref{fig3}(b)]. At the same time the populations have peaks when the quasi-levels approach each other [Fig.~\ref{fig3}(c)]. Also this effect occurs for interacting qubits.

%
%
%
\begin{figure}
\begin{center}
    \includegraphics[width=7cm,height=11cm]{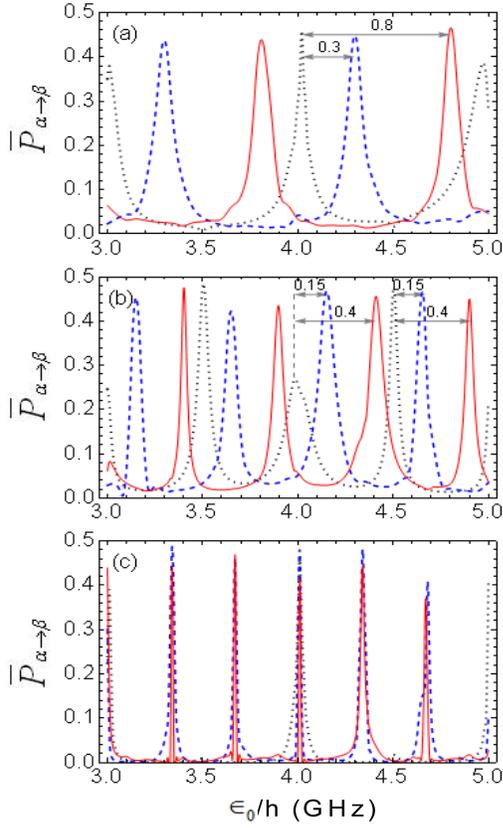}
\end{center}
\caption{\label{fig5}(Color online). The transition probabilities: $\overline{P}_{1\rightarrow2}$ (a), $\overline{P}_{1\rightarrow3}$ (b), and $\overline{P}_{1\rightarrow4}$ (c), as a function of the control parameter $\epsilon_{0}=\epsilon_{20} = 2\epsilon_{10}$ for different coupling constants $J$: black dotted lines $J=0$, dashed blue $J/h =-0.3$ GHz, and continuous red $J/h =-0.8$ GHz. Here we have set: $\Delta_{2}/h = 1.5 \Delta_{1}/h = 0.45$ GHz, $\omega/2\pi = 1$ GHz, and $A_{2}/h = 2A_{1}/h = 7$ GHz.}
\end{figure}
%
%
%
\begin{figure}[t]
\begin{center}
   \includegraphics[width=7cm,height=13cm]{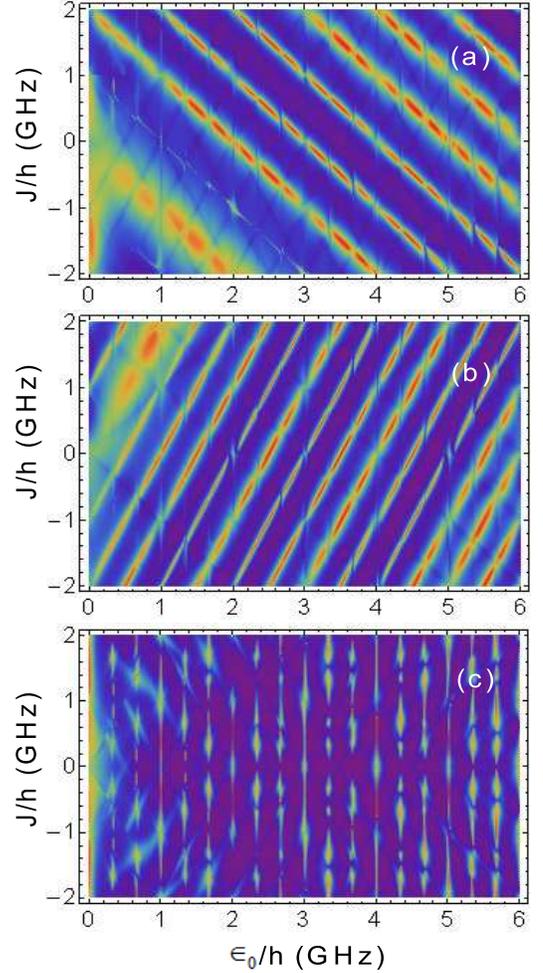}
\end{center}
\caption{\label{fig6}(Color online).  The transition probabilities: $\overline{P}_{1\rightarrow2}$ (a), $\overline{P}_{2\rightarrow4}$ (b), and $\overline{P}_{1\rightarrow4}$ (c) as functions of the control parameter $\epsilon_0 = \epsilon_{20} = 2 \epsilon_{10}$ and the coupling parameter $J$. The qubit parameters are the same as in Fig.~\ref{fig5} and the field amplitudes used here are $A_{2}/h= 2A_{1}/h = 7$ GHz. The color bar is the same as in Fig.~\ref{fig5}.}
\end{figure}
%
%
As can be seen from Fig.~\ref{fig3}(a, b), the quasi-energies exhibit a nontrivial dependence on the field amplitude for the two coupling parameters $J=0$ [see Fig.~\ref{fig3}(a)] and $J = -0.1$ GHz [see Fig.~\ref{fig3}(b)].
In this case, the appearance of additional quasi-energy crossings and the formation of new peaks for the transition probabilities might be possible here [see Fig.~\ref{fig3}(c, d)]. The circles show additional quasi-energy levels crossing and their coincidence with resonance peaks (the gray dashed lines in Fig.~\ref{fig3}).

The dependencies of the transition probabilities between the states of two qubits built at one time according to the alternating field amplitude and the control parameter are more informative and obvious. The interference patterns in Fig.~\ref{fig4} for the interacting qubits are qualitatively understandable on the basis of the results given in section II. The positions of the ``bright spots'' on the probability diagrams, at definite values of $\epsilon_{0}$ and the field amplitudes, coincide with the positions of the given transitions on the dependence of the quasi-energies on the amplitude, as shown in Fig.~\ref{fig3}. We see that the system possesses a distinct behavior depending on the coupling parameter $J$, which causes a shift of the peaks depending on $J$ along the bias direction. Figure~\ref{fig5} clarifies the radical change that the interaction between the qubits makes on the level populations. First, from the RWA analysis follows that a shift of the resonance peaks as a function of the coupling constant should be observed. These shifts can be seen in Fig.~\ref{fig5}(a) and (b) (in the transitions $\overline{P}_{1\rightarrow2}$ (a) and $\overline{P}_{1\rightarrow3}$ (b) the shifts with increasing the coupling constant of qubits are shown by the arrows).
Secondly, for the transitions $\overline{P}_{1\rightarrow4}$ the resonance peaks do not move when the coupling parameter $J$ is changed.

We note that in order to calculate the level populations for the coupled qubits in Fig.~\ref{fig5}, a definite relationship between the control parameters:
$\epsilon_{0}=\epsilon_{20}(t)=\lambda \epsilon_{10}(t)$ (where $\lambda$ is a parameter that determines the slope lines in the plane $\epsilon_{20}$ and $\epsilon_{10}$) has been assumed.
The analysis in the framework of the RWA (Section II) has shown that the locations of the resonance peaks are given by the following conditions:
\begin{eqnarray}
\epsilon_{20}+J+n\hbar\omega\approx0,\quad  ( 1\rightarrow2),\label{20} \\
\epsilon_{10}-J+n^{'}\hbar\omega\approx0, \quad (2\rightarrow4),\nonumber
\end{eqnarray}
and
\begin{eqnarray}
\epsilon_{10}+J+m\hbar\omega\approx0, \quad (1\rightarrow3),\label{21} \\
\epsilon_{20}-J+m^{'}\hbar\omega\approx0, \quad ( 3\rightarrow4).\nonumber
\end{eqnarray}
We can see in Figs.~\ref{fig2}, \ref{fig3} and Fig.~\ref{fig5} for the populations as well as for the interference patterns in Fig.~\ref{fig4}(b, d), that the resonance peaks  undergo a shift by a distance $| J |$ for the transitions $1 \rightarrow 2$ [see Fig.~\ref{fig4}(b) and Fig.~\ref{fig5}(a)]  and $3\rightarrow  4$. At the same time, for the transitions $1 \rightarrow  3$ [see Fig.~\ref{fig4}(d) and Fig.~\ref{fig5}(b)] and $2 \rightarrow 4$,  the peaks  are shifted by a distance $| J | / \lambda$.
Also shown in the figures is the fact that due to the chosen relations between the parameters (for example, when $\lambda=2$) and the relevant conditions ($\epsilon_{0}+J+n\hbar\omega\approx0$ and $2\epsilon_{0}+J+m\hbar\omega\approx0$) the  ``bright'' resonances of a quantum-coherent tunneling in the transition $1 \rightarrow 3$ [Fig.~\ref{fig4}(d)] are seen twice as often than for the transition $1\rightarrow 2$ [Fig.~\ref{fig4}(b)].
Depending on the sign of the coupling constant $J$ (ferromagnetic or antiferromagnetic coupling), there is a shift of the resonance peaks to the right or to the left.

Also note that the transitions to a higher excited level are due to virtual transitions that are possible when both of the paired resonance conditions Eq.~({\ref{20}}) and/or Eq.~({\ref{21}}) can be fulfilled with the participation of second and third intermediate levels, respectively.
A characteristic feature of this transition is the absence of peaks at integer values of the control parameter of the qubits, and the lack of resonance shifts  when the coupling constant $J$ is changed.
The positions of the resonance peaks (for $1 \rightarrow 4$) for fixed $J$ are determined by $  \epsilon_{0}=\frac{s\hbar\omega}{\lambda+1}$, where $s\equiv n+n^{'}=m+m^{'}$,  and do not depend on  the coupling constant.

Thus, Figs.~\ref{fig5}(a, b) demonstrate the shift of peaks when increasing the parameter $J$ and that agrees qualitatively with the results of the analysis on the basis of the RWA (see section II). These conclusions manifest the fact that the experimental study of the response of a system of coupled qubits will make it possible to obtain some additional information and in particular determine the qubit coupling parameter.

In Fig.~\ref{fig6}(a, b) we show the dependence of the population for the transitions $1 \rightarrow2$ and $2\rightarrow 4$ depending on the control parameter and the interaction parameters of qubits. For the selected slope parameter, $\lambda = 2$ is a clearly visible position of the resonance peaks, defined by Eq.~(\ref{20}).
Resonant lines defined by Eq.~(\ref{21}) look quite similar. In contrast, the resonance peaks in Fig.~\ref{fig6}(c) for the transition $1\rightarrow 4$ are determined by the intermediate states, so according to Eqs.~(\ref{20}) and Eq.~(\ref{21}) these will be located at the intersection of the lines.

\section{CONCLUSIONS}

In this work we have presented results on the behavior of two interacting qubits in a strong driving field. The principal difference of our approach from the works devoted to the laser spectroscopic investigations of multi-level atomic systems is that we study the excitation probability dependencies on the applied field amplitude and the control parameter at a fixed frequency of the applied field.

For a better understanding of the effects of driving fields on a multi-level system we use the RWA, which allows to find simple conditions of the system resonant excitation. We have shown that these conditions differ from those that occur in the case of a single qubit. The most important result here is that these conditions of the resonant excitation include the interaction qubit constant. The realized numerical simulation confirms the qualitative conclusions as follows from the RWA.

Our results show that the change of the field amplitude and the control parameter have a strong effect on the system dynamics. At the same time, the quasi-energy basis proves to be the most adequate for describing states in periodic time-dependent fields. The quasi-energy representation gives the precise intermediate states of a system in a driving field with an arbitrary amplitude and allows to detect the peculiarities of resonant transitions caused by quasi-level motion and crossing as a function of changing parameters. This numerical method of calculating quasi-energy states of multi-level systems made it possible to find the transition probabilities in a quasi-energy representation and build interference patterns for the transition probabilities.
The interference patterns obtained are very sensitive to the coupling strength of the qubits, suggesting a method to extract the value of the coupling parameter. The other parameters of the qubits, in particular the tunneling rates, also significantly affect the interference pattern so they can also be obtained in experiments.

The RWA as well as the numerical calculations of quasi-energy levels of the qubits in the strong driving field has shown that the effect of avoided crossings leads to drastically increased transition probabilities between the qubits steady states. Surprisingly, the peaks of the transition probabilities between the directly coupled states shift with changing the inter-qubit coupling $J$, but for indirectly coupled states the peak positions are not affected by $J$. This effect should be observed in experiments using the technique of amplitude spectroscopy.

The theory developed in this work should allow to extend the technique of amplitude spectroscopy used earlier for a single qubit \cite{Oliver,Berns1,Sillanpa,Berns2,Ruder} to more complicated systems. Clearly, amplitude spectroscopy can be used for studying the spectra of artificial quantum objects: quantum wells, quantum dots, quantum wires etc. in which the distances between energy levels are significantly smaller than in atomic systems.
\acknowledgments
We thank Sergey Shevchenko for useful comments on the manuscript.
SA and FN acknowledge partial support from the LPS, NSA, ARO, NSF grant No. 0726909, Grant-in-Aid
for Scientific Research (S), MEXT Kakenhi on Quantum
Cybernetics, and the JSPS-FIRST program. This work was funded in part by the Federal Program of the Russian Ministry of Education and Science through the contract No. 07.514.11.4012, and the RFBR Grants No. 11-02-97058-a and 12-07-00546-a. M.V.D.
was financially supported by the ``Dinastia'' fund.


\end{document}